\documentclass[preprint2]{aastex6}

\newcommand{\Ha}{H$\alpha$}

\newcommand{\Htwo}{H\textsubscript{2}}

\newcommand{\heI}{\ion{He}{1}}

\newcommand{\kms}{km~s{$^{-1}$}}

\newcommand{\cmsq}{cm$^{-2}$}

\newcommand{\Hplus}{H$^{+}$}
\newcommand{\Heplus}{He$^{+}$}

\newcommand{\tC}{{$\theta^1$~Ori~C}}
\newcommand{\tB}{{$\theta^1$~Ori~B}}
\newcommand{\tA}{{$\theta^2$~Ori~A}}

\newcommand{\hi}{\ion{H}{1}}

\newcommand{\oiii}{[\ion{O}{3}]}
\newcommand{\oii}{[\ion{O}{2}]}
\newcommand{\Cii}{[\ion{C}{2}]}
\newcommand{\oi}{[\ion{O}{1}]}

\newcommand{\nii}{[\ion{N}{2}]}

\newcommand{\siii}{[\ion{S}{3}]}
\newcommand{\Caii}{\ion{Ca}{2}}
\newcommand{\Nai}{\ion{Na}{1}}
\newcommand{\siiia}{\ion{S}{3}}
\newcommand{\Piii}{\ion{P}{3}}

\newcommand{\Vscat}{{\bf V$\rm_{scat}$}}
\newcommand{\Vscatnii}{{\bf V$\rm_{scat,[N II]}$}}
\newcommand{\Vscatoiii}{{\bf V$\rm_{scat,[O III]}$}}
\newcommand{\Vwide}{V$\rm_{wide}$}
\newcommand{\Vblue}{{\bf V$\rm_{blue}$}}
\newcommand{\Vbluenii}{{\bf V$\rm_{blue,[N II]}$}}
\newcommand{\Vblueoiii}{{\bf V$\rm_{blue,[O III]}$}}
\newcommand{\Vlow}{{\bf V$\rm _{low}$}}
\newcommand{\Vlowoi}{{\bf V$\rm _{low,[O I]}$}}
\newcommand{\Vlownii}{{\bf V$\rm _{low,[N II]}$}}
\newcommand{\Vlowoiii}{{\bf V$\rm _{low,[O III]}$}}
 \newcommand{\Vnewoiii}{{\bf V$\rm _{new,[O~III]}$}}
\newcommand{\Vnewnii}{{\bf V$\rm _{new,[N~II]}$}}
\newcommand{\Vnew}{{\bf V$\rm _{new}$}}

\newcommand{\Vwideoiii}{{\bf V$\rm_{wide,[O~III]}$}}
\newcommand{\Vwidenii}{{\bf V$\rm_{wide,[N~II]}$}}
\newcommand{\Vmifnii}{{\bf V$\rm_{mif,[N~II]}$}}
\newcommand{\Vmifoiii}{{\bf V$\rm_{mif,[O~III]}$}}
\newcommand{\Vmif}{{\bf V$\rm_{mif}$}}

\begin{document}

\title{The structure of the Orion Nebula in the direction of \tC}

\author{N. P. Abel\affil{1}}
\affil{MCGP Department, University of Cincinnati, Clermont College, Batavia, OH, 45103}

\author{G. J. Ferland\affil{2}}
\affil{Department of Physics and Astronomy, University of Kentucky, Lexington, KY 40506}

\and

\author{C. R. O'Dell\affil{3}}
\affil{Department of Physics and Astronomy, Vanderbilt University, Nashville, TN 37235-1807}

\begin{abstract}

We have used existing optical emission and absorption lines, \Cii\ emission lines, and \hi\ absorption lines to create a new model for a Central Column of material near the Trapezium region of the Orion Nebula.
This was necessary because recent high spectral resolution spectra of optical emission lines and imaging spectra in the \Cii\ 158 $\mu$m line 
have shown that there are new velocity systems associated with the foreground Veil and the material lying between \tC\ and the Main Ionization Front of the nebula. When a 
family of models generated with the spectral synthesis code Cloudy were compared with the surface brightness of the emission lines and strengths of the Veil absorption lines seen in the Trapezium stars, distances from \tC, were derived,
with the closest, highest ionization layer being 1.3 pc. The line of sight distance of this layer is comparable with the size of the inner Huygens Region in the plane of the sky.
These layers are all blueshifted with respect to the Orion Nebula Cluster of stars, probably because
of the pressure of a hot central bubble created by \tC 's stellar wind. We find velocity components that
are ascribed to both sides of this bubble. Our analysis shows that the foreground \Cii\ 158 $\mu$m emission is part of a previously identified layer that forms a portion of a recently discovered expanding shell of material covering most of the larger Extended Orion Nebula.
\end{abstract}


\section{Introduction}
\label{sec:Intro}
As the closest star formation region that has a bright H II region, the Orion Nebula presents a unique
testing-ground for theories of star formation and the physical processes governing astronomical plasmas.
This opportunity has created a rich literature on these subjects \citep{fer01,ode01,mue08,ode08,goi15,kong}. This article 
reports on a detailed study of material beyond and in the foreground of the dominant ionizing star (\tC) in the direction of the Trapezium stars. The brightest components (we shall call them layers as we study only a small area in the plane of
the sky) lie beyond \tC\ and are largely associated with emission from the blister of ionized gas on the surface of the host molecular cloud. Less visible layers lie in the foreground of \tC\ and are collectively called the Veil. The Huygens Region 
falls near the northeast corner of a much lower surface brightness elliptical bubble known as the Extended Orion Nebula (EON, \citet{gud08}).

\subsection{Background of this study}
\label{sec:Background}

Orion's Veil is a series of atomic and ionized layers, physically associated with the Orion star-forming environment.  The Veil covers the brightest part of M42 (the Huygens Region) and is the primary source of extinction for the Nebula and for the stars ionizing M42, the Trapezium cluster. \citet{ode00} have mapped the extinction across the entire face of the Nebula, with about 1.6 mag of extinction towards the Trapezium.  Additionally, \citet{vdw90}, Troland, Heiles, \& Goss (1989) and \citet{tom16} have mapped the magnetic field strength in the Veil through \hi\ 21~cm absorption.  

The simultaneous mapping of extinction and magnetic field for an environment with a well-characterized spectral energy distribution (SED) has made the Veil an ideal testbed to understand the physics of the interstellar medium (ISM), in particular the partitioning of magnetic, turbulent, and thermal energetics.  In a series of works (\citet{abel04}; \citet{abel06}; \citet{abel16}) combined UV and optical absorption line data towards the Trapezium with the existing magnetic field and extinction maps to produce detailed calculations of the physical conditions in the two primary atomic layers seen in \hi\  21~cm absorption towards the Trapezium (typically referred to as Components A and B).  Utilizing the spectral synthesis code Cloudy \citet{fer17}, these works computed the density, temperature (gas and spin), and chemical composition for each component, then used this data to compare magnetic, thermal, and turbulent energies.  Overall, the combination of observations and theoretical modeling showed Component B (2.0 pc) to be closer to the Trapezium than Component A (4.2 pc), with Component B also being somewhat warmer and denser than Component A.  Finally, both layers appeared to be magnetically dominated, meaning the energy associated with the magnetic field of each component exceeds both the thermal and turbulent energies. 

Although the series of Abel et al. papers focused primarily in the atomic Veil layers, an H\textsuperscript{+} region must exist between the Trapezium and Component B.  In the optical, 2S\textsuperscript{3 }He\textsuperscript{0} absorption at 388.9 nm is observed towards the Trapezium and \tA.  In the UV,  S\textsuperscript{2+} and P\textsuperscript{2+} absorption is observed towards the Trapezium \citet{abel16}.  While the series of Abel et al. papers attempted to account for the ionized gas in their modeling efforts, the lack of emission features associated with this ionized component (due to the much brighter M42 on the other side of the Trapezium) have made it difficult to state anything definitive.

Two recent studies allow the possibility of understanding the ionized component between the Trapezium and Component B.  \citet{ode18} performed a detailed study of the velocity components from the inner portions of the Orion Nebula and found optical [N II] (658.4 nm) and [O III] (500.7 nm) emission features at similar velocities to the 2S\textsuperscript{3 }He\textsuperscript{0}, S\textsuperscript{2+}, and P\textsuperscript{2+} absorption features seen in the optical and UV.  These emission features are critical, as they allow the calculation of the physical properties of the ionized component, which can then be used to determine the effects of the ionized component model on the model results for the atomic layers.  Another recent study, by \citet{ode17}, calculated the relative effects of \tC\ and another nearby O star, \tA, on the ionization balance throughout the Nebula.  This study found, in many regions of Orion, that \tA\ cannot be ignored as a source of ionizing photons.  The effects of \tA\, therefore, should be explored in any calculation of the ionized component.  Additionally, in the light of \citet{ode17}, the degree to which \tA\ affects the neutral layers along the line of sight also requires investigation.

\subsection{Outline of this paper}

The format of this work is as follows.  In Section 2, we highlight the key observations used in this study, emphasizing the recent work of \citet{ode18} along with previous optical, UV, and radio studies of the region and the line of sight towards the Trapezium.  Section 3 presents revised models of the Veil, taking into account the \citet{ode18} emission-line observations of [N II], [O III], and [O I] along the line of sight.  Section 4 explores the effects of including the SED of \tA\ on the model results, while Section 5 discusses the velocity systems in the Orion environment and the underlying physical mechanisms driving the observed velocity correlations.  We end with a series of conclusions in Section 6.

Throughout this paper we designate radial velocities in the the Heliocentric system. These may be converted to LSR velocities by subtracting 18.1 \kms. We adopt the distance of 383$\pm$3 pc derived by \citet{mk17}, which is in agreement with more recent results using Gaia DR2 \citep{gro19}. This distance 
gives a scale of 1.86$\times$10$^{-3}$ pc/\arcsec.

\section{Observations} \label{sec:obs}
\label{sec:observations}

We draw on previously published optical window observations, particularly those in the high spectral resolution Spectroscopic Atlas of Orion Spectra \citep{gar08} (henceforth `the Atlas') prepared with spectra from Kitt Peak National Observatory, Cerro-Tololo Interamerican Observatory, and San Pedro M\'artir Observatory. The Atlas was composed from north-south slit spectra at intervals of 2\arcsec\ in Right Ascension and have a velocity resolution of 10 \kms. The spatial resolution in Declination was seeing limited at about 2\arcsec. The \oiii\ observations of \citet{doi04} were used in creating the Atlas and a good illustration of a spectrum is given in Fig. 1 of that study. We also used the lower spectral resolution study made with the MUSE 2--D spectrograph at the European Southern Observatory's Paranal observatory \citep{wei15}. In this case the velocity resolution was about 107 \kms\ and the spatial resolution was about 1\arcsec.
        
 In the ultraviolet we employed the HST ultraviolet spectra of \tB\ at 2.5 \kms\ resolution described in detail in \citet{abel16}. All of the lines of interest were seen in absorption against the stellar continuum of \tB.

We also used the results from \cite{goi15}. That study included all of the Huygens Region but did not extensively go into the EON. It reported Herschel space craft spectra of the \Cii\ 158 $\mu$m line at 0.2 \kms\ and 11\farcs4\ resolution. The study also presented H41$\alpha$ observations with the IRAM-30m telescope. These had 0.65 \kms\ and 27\arcsec\ resolution. Their discussion also used CO 2--1 observations by \citet{ber14} at 0.4 \kms\ and 11\arcsec\ resolution. 
        
We also employed calibrated emission line images made with the Hubble Space Telescope \citep{ode96,ode09} and lower resolution ground-based camera images by Robert Gendler reproduced in \citet{ode10}. The latter cover the entire EON region and go beyond the mosaic of HST images first reproduced in \citet{hen07}.

\subsection{Surface Brightnesses and column densities of lines used in our study}
\label{sec:SurfBrt}

Our analysis has required derivation of the extinction corrected surface brightness of the \Vlow\ velocity component discussed in \citet{ode18}.
A refined definition of the multiple velocity components is given in Table~\ref{table:criteria} and discussed in Section~\ref{sec:VelSys}. 

In the case of \nii\ 658.4 nm and \oiii\ 500.7 nm the spectra were averaged over a 3200 arcsec$^{2}$  square centered on the Trapezium stars. The derived surface brightness values were corrected for interstellar extinction using c$_{H\beta}$~= 0.6 from \citet{ode00} and the 
reddening curve of \citet{bla07}.
The results were S(\Vlownii)~=~2.8~$\pm$1.4~$\times$10$\rm ^{-3}$ ergs s$^{-1}$ cm$^{-2}$ sr$^{-1}$
and S(\Vlowoiii)~=~1.1~$\pm$0.7~$\times$10$\rm ^{-2}$ ergs s$^{-1}$ cm$^{-2}$ sr$^{-1}$.

\citet{ode18} established that a \Vlow\ component of the \oi\ 630.0 nm line could be seen in high velocity resolution spectra obtained when the Earth's motion has shifted the night sky \oi\ sufficiently to deconvolve it from the nebular emission.  This was the case for three slit positions in the Atlas, one in the \oi\ velocity study of \citet{ode92a}, and it also appears in the high resolution spectrum of \citet{jack00}. Unfortunately none of these detections
fall within the region used to obtain the surface brightnesses in \nii\ and \oiii. However, they lie close to this region and the ratio of the signals of 
the \Vlow\ and V$\rm_{mif}$ components are quite similar, at 0.10$\pm$0.01. We then used the extinction--corrected images in the MUSE study \citep{wei15} to 
determine the surface brightness of the Main Ionization Front (MIF) component and then scaled this down by a factor of ten to find 
S(\Vlow ,\oi)~=~1.6~$\pm$0.53~$\times$10$\rm ^{-4}$ergs s$^{-1}$ cm$^{-2}$ sr$^{-1}$.

The final line used in the analysis was the 388.9 nm line of He~I. Most of the helium lies in the metastable 2$^{3}$S state that is populated as ionized helium recombines. In the
study of absorption lines in Trapezium stars \citep{ode93} a column density of 1.45$\times$10$^{13}$  ions cm$^{-2}$ was found in \tC.

\subsection{Characteristic Velocity Systems}
\label{sec:VelSys}

\begin{figure}
  \plotone{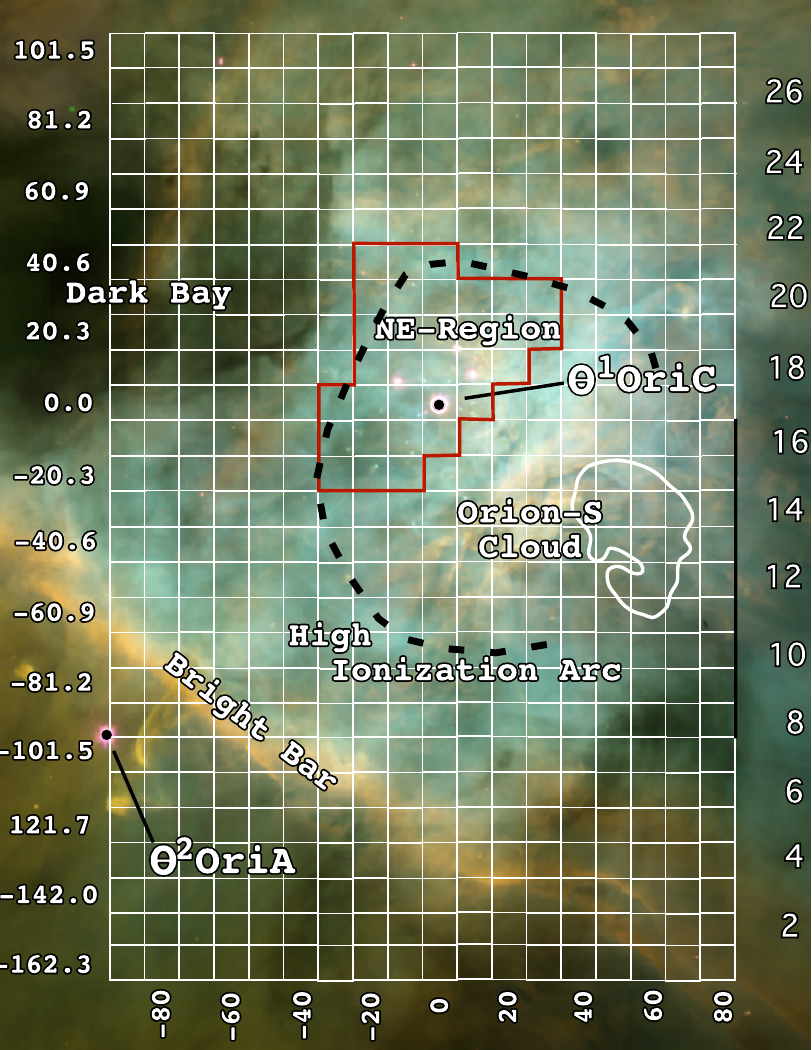}
    \caption{This 233\arcsec $\times$302\arcsec\ image of a portion of the Huygens Region \citep{ode96} is color coded: Blue  \oiii , green \Ha , and red  \nii\ emission. North is up, west is to the right. The left hand labels indicate the distance of the center of each sample from \tC\ in arcseconds. The right hand labels indicate the declination samples in the numbering system used in \citet{ode18}. The bottom
 labels indicate the east-west distance from \tC , again in arcseconds. The red lines outline the NE--Region and the black dashed line the High Ionization Arc.}
\label{fig:HRone}
\end{figure}

In the study of the brightest part of the the Huygens Region \citet{ode18} averaged the Atlas spectra over areas of 10\arcsec$\times$10\arcsec , thus increasing the signal to noise ratio. This grid is shown in Figure~\ref{fig:HRone}.
In order to identify systematic patterns in the \nii\ and \oiii\ velocity components we defined an area within this grid of Atlas spectra, as shown in Figure~\ref{fig:HRone}, designated as the NE Region. It was selected to include the dominant ionizing star \tC\ and to be free of major shocks and jets. 

\begin{deluxetable*}{lll}
\tablenum{1}
 \tablecaption{Criteria for Identifying Velocity Components  \label{table:criteria}}
\tablehead{
\colhead{Ion} &
\colhead{Component} &
\colhead{Criteria}
}
\startdata
 \nii & \Vscat & Longer of two red components or the single red component if S(\nii)/S(mif,\nii)$\leq$0.1.\\
 \nii & \Vnewnii  & Shorter of two red components or S(obs,\nii)/S(mif,\nii)$\geq$0.1 when there is one red component\\
\nii& \Vmifnii &The strongest component with FWHM$\leq$18.00 \kms .\\
 \nii &\Vwidenii &  A \Vmifnii\ component with FWHM$\geq$18.00 \kms .\\
\nii &\Vlow & The longer of two blue components or S(obs,\nii)/S(mif,\nii)$\geq$0.05 when there is one\\
 \nii  &\Vblue\ & Shorter of two blue components when there are two or S(\nii)/S(mif,\nii)$\leq$0.05 when there is one \\
 \oiii & \Vscat & Longer of two red components or the single red component if S(\oiii)/S(mif,\oiii)$\leq$0.1.\\
\oiii & \Vnewoiii  & Shorter of two red components or S(obs,\oiii)/S(mif,\oiii)$\geq$0.1 when there is one red component.\\
 \oiii & \Vmifoiii & The strongest component with FWHM$\leq$16.00 \kms .\\
\oiii &\Vwideoiii &  A \Vmifoiii\ component with  FWHM $\geq$16.00 \kms .\\
\oiii &\Vlow &The longer of two blue components or S(\oiii)/S(mif,\oiii)$\geq$0.07 when there is one.\\
\oiii  &\Vblue\ & Shorter of two blue components or S(\oiii)/S(mif,\oiii)$\leq$0.07 when there is one \\
\enddata
\end{deluxetable*}

  \begin{figure}
	\includegraphics
	[width=\columnwidth]
	{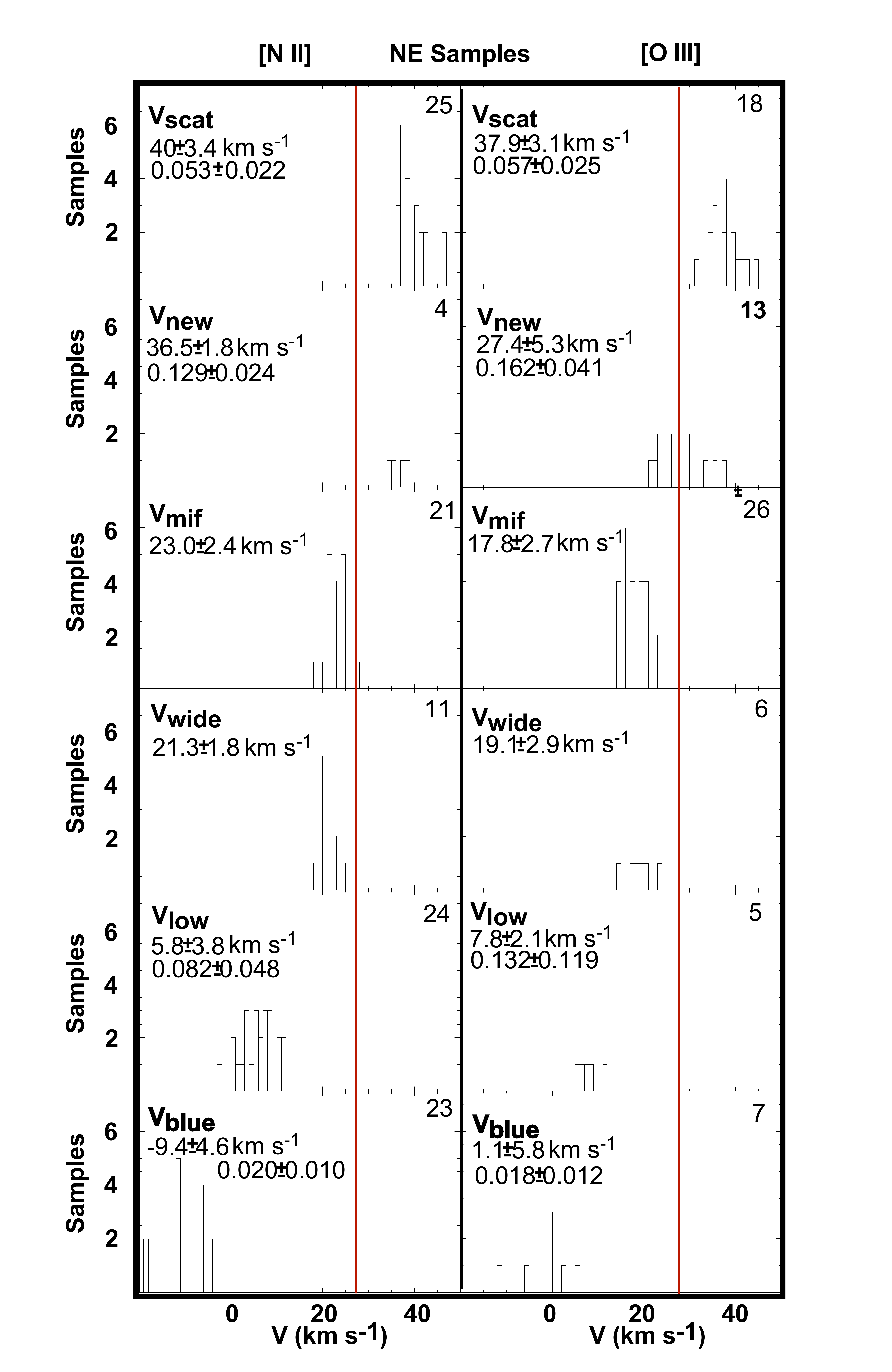}
    \caption{Histograms of the velocity components within the NE--Region are shown here in descending order of average velocity and discussed in Section~\ref{sec:VelSys}.
    The average velocity, average ratio of S(component)/S(mif), and the total number of samples are shown for each component. The red line indicates the average velocity of the PDR (27.3 \kms ).}
\label{fig:Histograms}
\end{figure}

We have examined each of the 10\arcsec$\times$10\arcsec averaged spectra within the NE--Region, assigning velocity components according to the criteria given in Table~\ref{table:criteria}. A good illustration of one of these \nii\ spectra and its deconvolution is shown in Fig.~A1 of \citet{ode18}. Sometimes there were two components to the red or blue of the \Vmif\ or \Vwide\ components, making the assignments easy. In the case of there being only one, then the signal of that component as compared with the corresponding \Vmif\ or \Vwide\ was used for assignment. These criteria produced an almost unambiguous assignment of each component. The results are shown in Figure~\ref{fig:Histograms}, with the assigned errors being based on the scatter of the values. The accuracy of very weak component lying on the shoulder of the strong \Vmif\ or \Vwide\ component is discussed in the Appendix of \citet{ode18}.  In this section the velocity groups are arranged in order of descending radial velocities, whereas in Section~\ref{sec:what} and Table~\ref{table:velocities} they are arranged according to their probable distance from the observer. All of these optical window lines arise from the MIF which is composed of material photo-evaporating towards the observer from the Photon Dominated Region (PDR).
The PDR lies immediately beyond the MIF and is a shock compressed region lying along the surface of the Orion Molecular Cloud (OMC).

The {\bf \Vscat} components in the Huygens Region are usually attributed to backscattering by dust in the PDR. The expected redshift of this component, relative to the 
layer producing the \Vmif\ emission is twice the photo-evaporation velocity. The samples to the east of the Trapezium and close to the Orion-S Cloud  in \citet{ode18} give velocity differences of 17 \kms\ for \nii\ and 19 \kms\ for \oiii , both consistent with expected evaporation velocities. Our NE--Region gives a similar velocity difference of 17$\pm$3 \kms\ for \nii\ and 20$\pm$3 \kms .
In the NE--Region, the \Vscatnii\ component appears in 25 of 32 samples and \Vscatoiii\ in 18.

The {\bf \Vnew} components were originally reported in \citet{ode18}, where they were only detected in \oiii. We now find a few \Vnewnii\ components
in the NE--Region (4 of 32 samples), but more (13 of 32 samples) in \Vnewoiii .
The larger values of the velocities for \nii\ (37$\pm$2 \kms ) than in \oiii\ 
(27$\pm$5 \kms ) argues against the interpretation in \citet{ode18} that the \Vnew\ component is photo-evaporation from a foreground component of the Veil.

The average of the {\bf \Vmif} and {\bf \Vwide} components provides a good measure of the velocity in the densest layers along the observer's line-of-sight.  In the NE--Region this is 22$\pm$3 \kms\ for \nii\ and 18$\pm$3 \kms\ for \oiii. These values are consistent with their being accelerating
photo-evaporation flow away from the Photon-Dominated Region (PDR) at 27.3 \kms\ \citet{ode18}. 

The {\bf \Vlow} emission line velocities have been interpreted 
\citep{abel16,ode18} as arising from ionized gas lying on the far side of the Veil component closest to \tC.  \Vlownii\ (6$\pm$4 \kms) and \Vlowoiii (8$\pm$2 \kms) velocities in the NE-Region fit with this interpretation and support the basis of our ionization modeling of the Veil. Although these lines are weak and lie on the blue shoulder of the MIF emission, \citet{ode18} tested their reality by creating artificial spectra with blue shoulder lines of varying strength and displacement. He concluded that the \Vlow\ components are real. The agreement of the Component I emission and absorption line velocities gives further confimation
of the ionized Component I as real. 

The {\bf \Vblue} components are the most enigmatic, with no current explanation. They are weaker than the \Vlow\ components, even after factoring in the classification criteria.
The velocities are more negative than the \Vlow\ components and have a greater range. They are most common (23 of 32 samples) in the NE-Region \nii\ spectra.

\section{Revised Theoretical Calculations}
\label{sec:calc}

\subsection{Computational Details }
\label{sec:comp}
We use c17 of the spectral synthesis code Cloudy \citep{fer17} to find a self-consistent model for each component observed along the line of sight to the Trapezium.  The assumed gas and dust abundances,infrared, optical, and UV stellar continuum (SED), and grain properties are all identical to \citet{abel16}, which in turn follows the general logic of the previous modeling efforts of the Veil presented in \citep{abel04,abel06} and \citet{lyk10}.  The primary difference between the calculations presented here and the one presented in previous works is the modeling of two ionized components along the line of sight (which must lie between the components observed in 21 cm absorption and the Trapezium) and the way we propagate the radiation field to model subsequent components along the line of sight.\  These components are seen in Figure~\ref{fig:Geometry} and are described below:

~~I. The ionized gas component closest to the Trapezium, with observed values for the surface brightness of \nii\ 658.3 nm and \oiii\ 500.7 nm.  In addition, the 2S\textsuperscript{3 }He\textsuperscript{0} (henceforth He I*) S\textsuperscript{2+} and P\textsuperscript{2+} column densities are known from previous absorption studies in the optical and UV {\citep{ode93,abel06}.  \par

II.  An ionized gas component observed in \oi\ 630.0 nm emission lying between the \nii\ and \oiii\ emitting region and the \hi\ 21 cm components.  \par

III. The \hi\ 21 cm component designated as Component B in previous Veil studies, whose velocity is associated with excited state H\textsubscript{2}\ observed in UV absorption.  Column densities for a variety of elements are derived in \citet{abel06}, and is the source of about two-thirds of the total \hi\ column density along the line of sight. For clarity this component will frequently be designated as III (B).\par

IV.  The \hi\ 21 cm component, previously designated as Component A, with similar column densities observed in this component to that of Component III, with the exception being the lack of molecular hydrogen.  The remaining one-third of the total \hi\ column density lies in this component. For clarity this component will frequently be designated as IV (A).\par

\vspace{\baselineskip}
 \begin{figure}
	\includegraphics
	[width=\columnwidth]
	{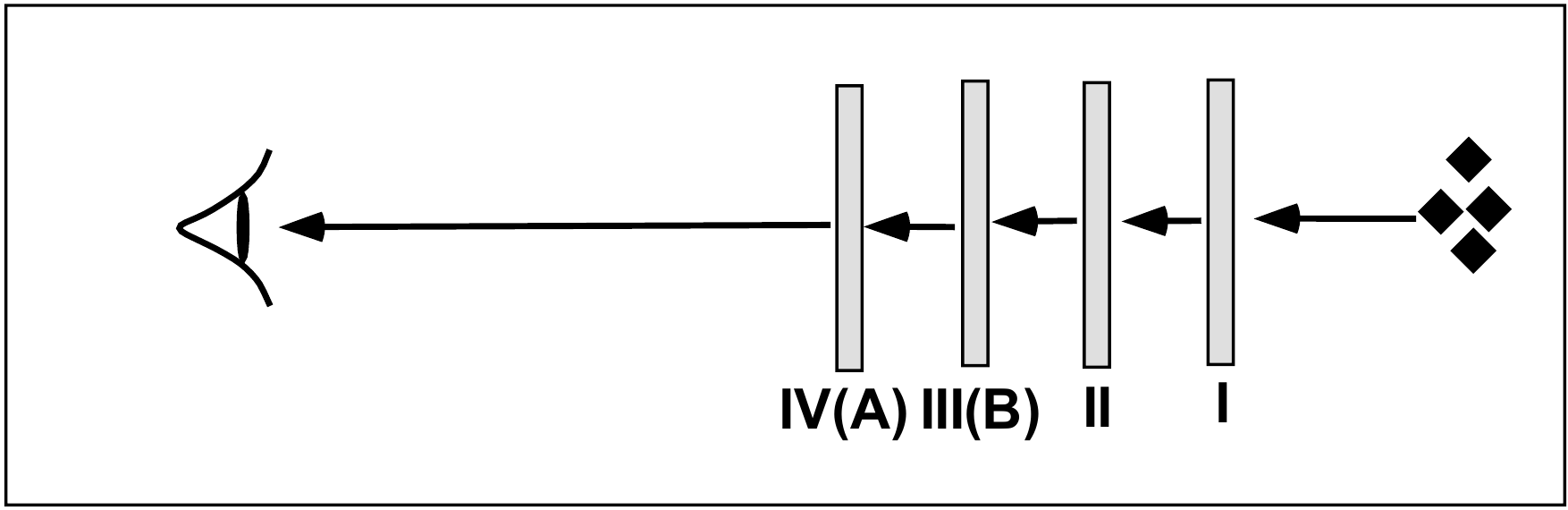}
    \caption{Assumed Model Geometry - Roman numerals correspond to the regions outlined in Section~\ref{sec:comp}. The four Trapezium stars are on the right and the observer on the left.}
\label{fig:Geometry}
\end{figure}

The velocities of the individual components make it likely that none of the four components listed are physically\ adjacent\ to each other.  However, since all components are affected by the same SED, each component is coupled through radiative transfer.  The radiation which makes it through the first component will be the incident continuum impingent upon the 2\textsuperscript{nd}\ component,\ which continues until the radiation field which makes it through \hi\ 21 cm component A, which ends up heading towards the observer.  Our calculations account for this effect.  Once we find the best-fitting model which accounts for the observed \nii , \oiii , and 2S\textsuperscript{3} He\textsuperscript{0}\ column\ density, we save the transmitted SED for the last zone of the calculation, and use this transmitted SED as the incident SED for the next component.  For each component, we vary the distance to the Trapezium and the total hydrogen density of the layer (assuming constant density, where "total" is defined as the sum of the hydrogen atoms per cm$^{-3}$ in ionized, atomic, and molecular form).  For component I, we stop all calculations at the observed \textit{N}(S\textsuperscript{2+}) column density of 10\textsuperscript{15} cm\textsuperscript{-2}.\  For component II, each calculation went until the H\textsuperscript{+}/H\textsuperscript{0} ionization fraction reached 10\textsuperscript{-2}.\  Finally, for components III (B) and IV (A), we stopped the calculation at the deduced H\textsuperscript{0} column density of 10\textsuperscript{21.5} cm\textsuperscript{-2} (component B) and 10\textsuperscript{21.2} cm\textsuperscript{-2}\ (component A).  
\subsection{Component I}
\label{sec:compI}

Figures~\ref{fig:NII} and \ref{fig:OIII} show the summary for the modeling of the \nii\ and \oiii\ region and the regions where the observations agree with these models.  Figure~\ref{fig:HeI*} is a similar figure for 2S\textsuperscript{3} He\textsuperscript{0}\ column density, where  we assume an error bar of $ \pm $ 0.3 dex since \citet{ode93} only report the average column density of 10\textsuperscript{13.3} cm\textsuperscript{-2}. The best fit to the observations (shown by the dot in Figure 6) yield S([N II]) = 3.97x10\textsuperscript{-3}ergs s$^{-1}$ cm$^{-2}$ sr$^{-1}$,S([O III]) = 1.6x10\textsuperscript{-2}ergs s$^{-1}$ cm$^{-2}$ sr$^{-1}$, and a HeI* column density of 10\textsuperscript{12.98} cm\textsuperscript{-2}, 
 corresponding to a density of around 10\textsuperscript{2.65 }cm\textsuperscript{-3}, and a distance of 10\textsuperscript{18.6}\ cm~ (1.3 pc)}.
 \begin{figure}
	\includegraphics
	[width=\columnwidth]
	{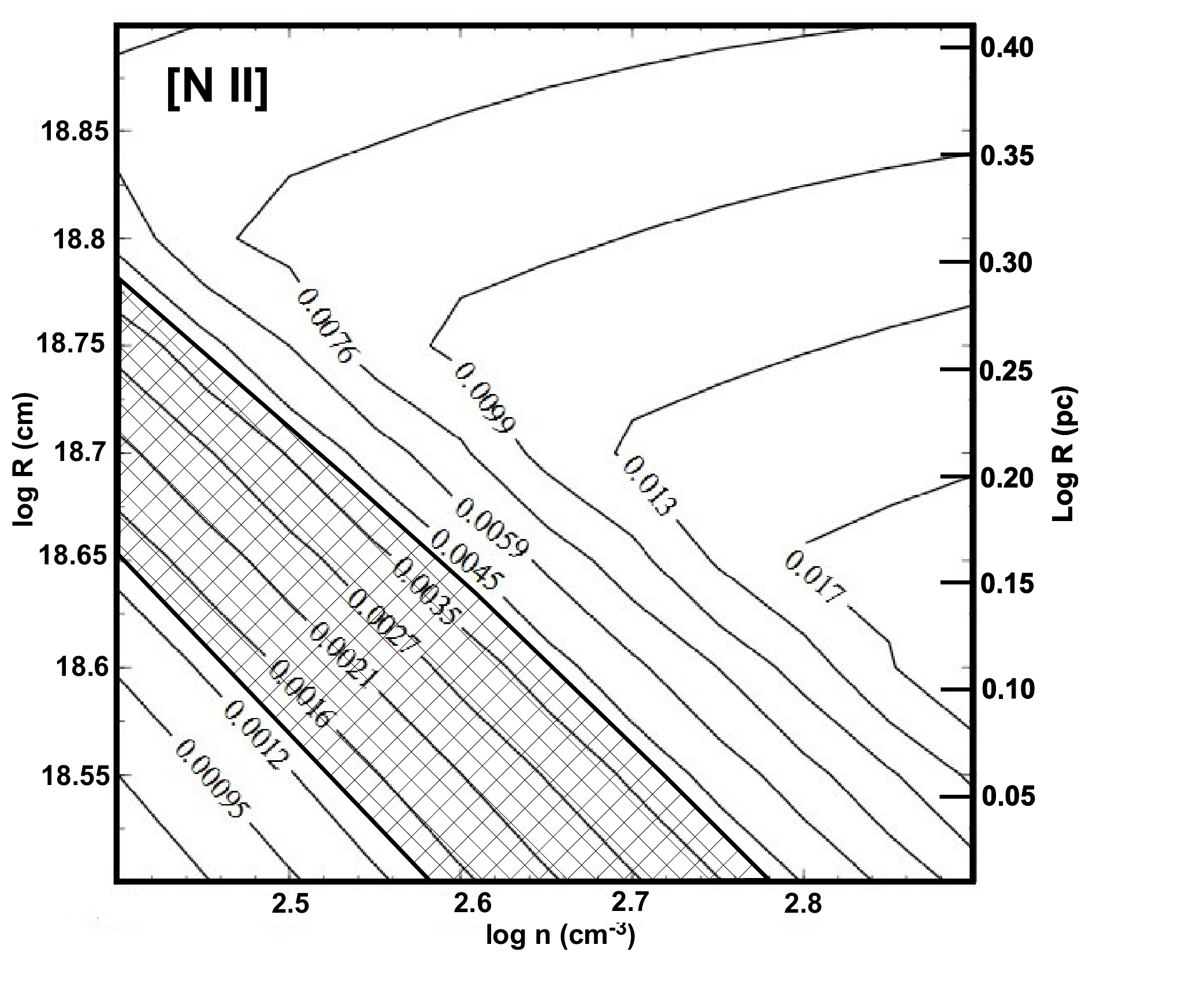}
 \caption{\nii\ 658.3nm surface brightness as a function of density and distance from the Trapezium for Component I. Lines of constant  predicted surface brightness in ergs s$^{-1}$ cm$^{-2}$ sr$^{-1}$ are labeled.}
\label{fig:NII}
\end{figure}
\vspace{\baselineskip}
\begin{figure}
	\includegraphics
	[width=\columnwidth]
	{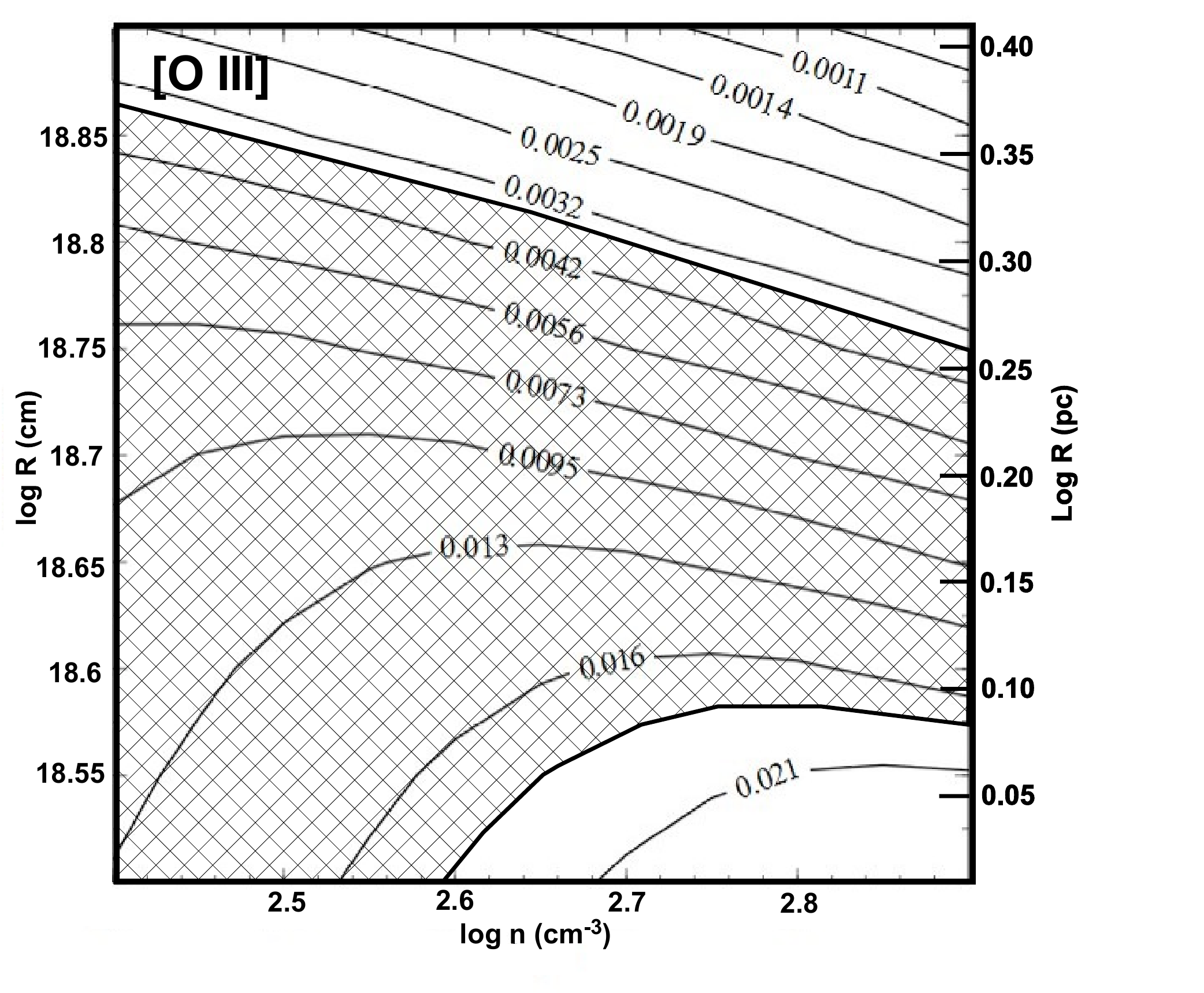}
    \caption{Like Figure~\ref{fig:NII} except now for \oiii 500.7 nm.}
\label{fig:OIII}
\end{figure}
\vspace{\baselineskip}
\begin{figure}
	\includegraphics
	[width=\columnwidth]
	{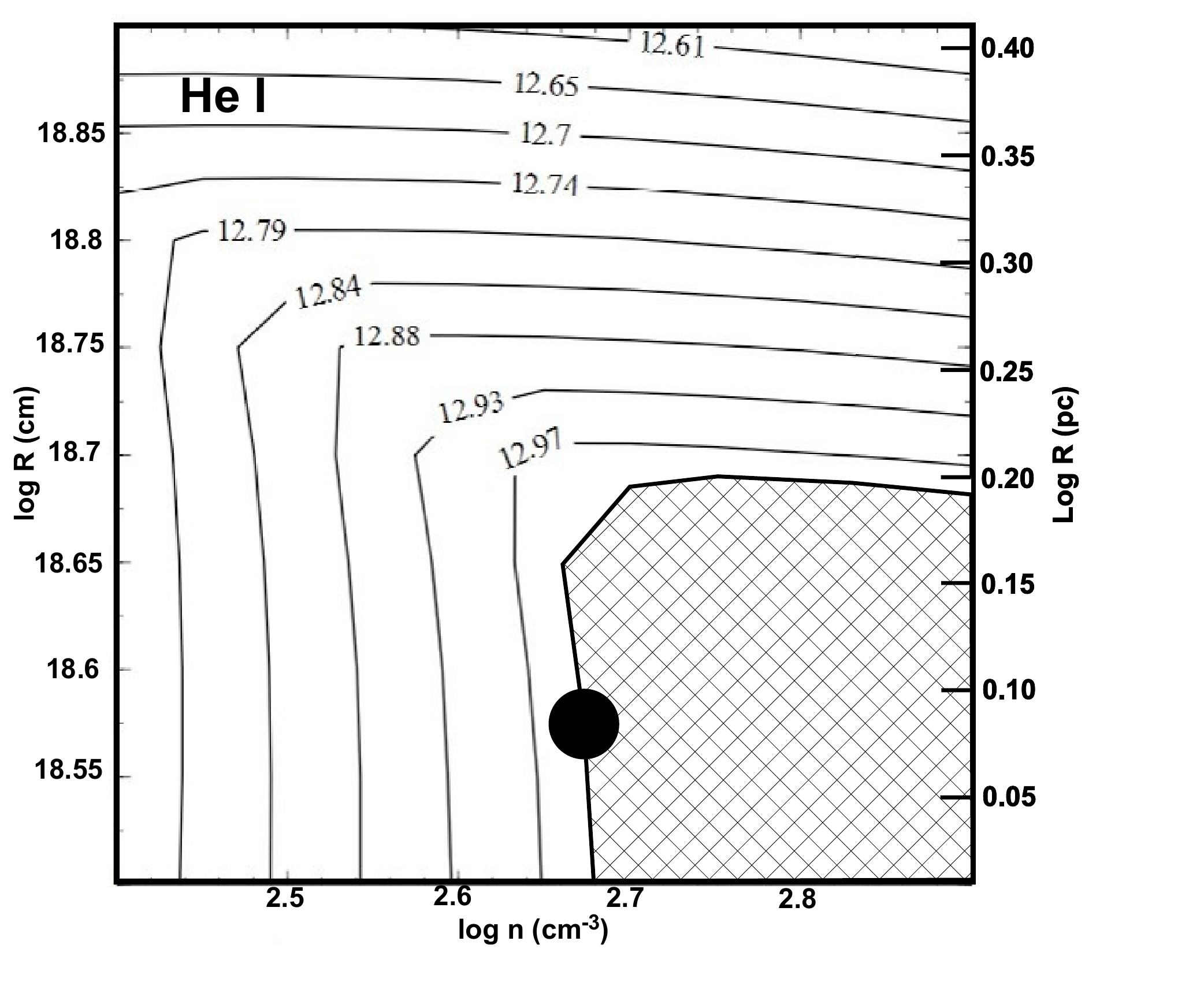}
    \caption{Like Figures~\ref{fig:NII} and \ref{fig:OIII} except now for values of the HeI* column density in Component I. The filled circle indicates the small region where the \nii , \oiii , and HeI* results agree.}
\label{fig:HeI*}
\end{figure}
\subsection{Component II}
\label{sec:CompII}

\begin{figure}
	\includegraphics
	[width=\columnwidth]
	{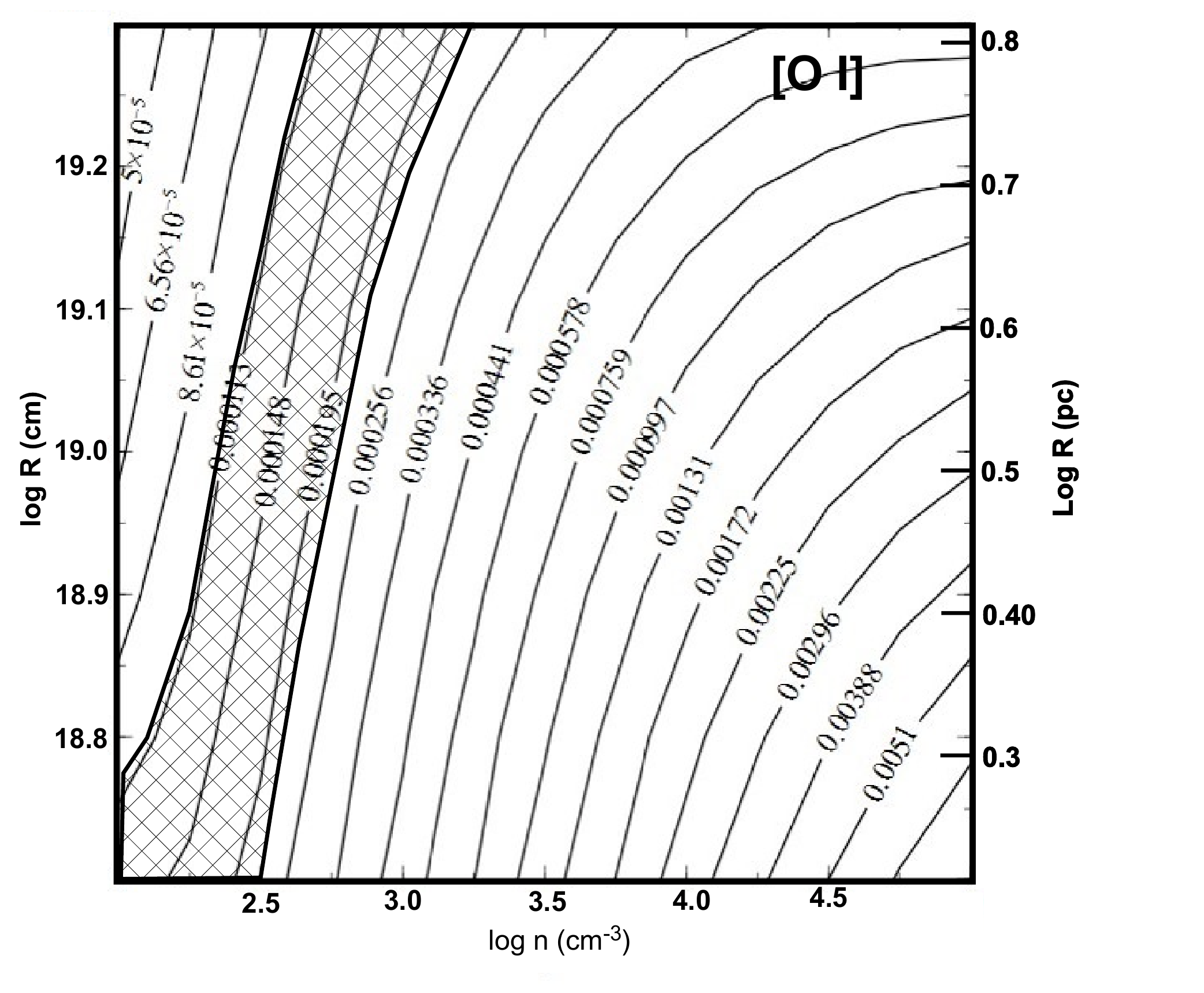}
    \caption{Predicted \oi\ 630.0 nm surface brightness as a function of density and distance from the Trapezium for Component II}
\label{fig:OI}
\end{figure}

Given the density/distance combination deduced for Component I, we saved the transmitted continuum for the last zone of this calculation and used this as the input SED to model Component II, which represents the \oi\ emitting region. Calculations for this component spanned a distance of log R (cm) = 18.7 -- 19.3, in increments of 0.1 dex, and a density range of log \textit{n} (cm$^{-3}$) = 2 -- 5 .  The lower range of distance was chosen so that it is farther from the Trapezium than component I.  Figure~\ref{fig:OI} shows the overall results, that is, both the calculated values and the range that agree with the observations.  A broad range of distances exists where the \oi\ observations are reproduced.  However, we can exclude some of this parameter space as unphysical, as the \oi\ region cannot be too far away from the Trapezium. \citet{abel06} showed that if the neutral Veil layers are much farther than 10\textsuperscript{19} cm from the Trapezium, then significantly more molecular hydrogen will form, contrary to the low H\textsubscript{2}\ abundance observed in the Veil.  Therefore, we have set as an upper limit for the \oi\ distance as 10\textsuperscript{18.9} cm, so that in the modeling of component III (B), the closest initial distance of to the Trapezium would not exceed 10\textsuperscript{19} cm.\  This leads to three distance/density combinations for the \oi\ emitting layer (at a resolution of 0.25 dex in density, 0.1 dex in distance):\par

	~~~~~I.   log \textit{n}  (cm$^{-3}$) = 2; log \textit{R (cm)} = 18.7 \par
	~~~~II.  log \textit{n}  (cm$^{-3}$) = 2.25; log \textit{R (cm)} = 18.7 -- 18.9\par
	~~~III. log \textit{n}  (cm$^{-3}$) = 2.5; log \textit{R (cm)} = 18.8 -- 18.9

For each of these combinations, we saved the transmitted continuum for use in modeling Component III (B) in Section~\ref{sec:compIII}.

\subsection{Component III (B)}
\label{sec:compIII}
Using the transmitted SED through the \oi\ emitting layer, we computed the predicted SED for component III (B).\   As with the first two components, we varied the distance and density away from the Trapezium, with the closest distance being 0.1 dex farther than the distances listed for the six combinations in Section~\ref{sec:CompII}.  The upper limit to the distance for all calculations is 10\textsuperscript{19.3} cm, which assures all calculations go beyond the upper limit to the total H\textsubscript{2} column density of 10\textsuperscript{17.35} cm\textsuperscript{-2}.\  The density was varied from log \textit{n}  (cm$^{-3}$) = 2 -- 4, in increments of 0.2 dex, which covers about 1 dex on either side of the density derived in \citet{abel06} for Component B (log \textit{n}  (cm$^{-3}$) = 3.1).\  All calculations, just as in  \citet{abel04} are performed until an H\textsuperscript{0} column density of 10\textsuperscript{21.5 }cm\textsuperscript{-2}\ is reached.  \par

The primary results of these calculations show the best model for component III (B) depended only on the density and distance of the Veil from the Trapezium, and not on the density-distance of the \oi\ emitting layer.  The predicted CI, CI$\ast$ , CI$\ast$ $\ast$ , and H\textsubscript{2} column density for Component III (B) was identical for the input SED corresponding to the \oi\ layer of log \textit{n} (cm$^{-3}$) = 2; log \textit{R (cm)} = 18.7 and the log \textit{n}  (cm$^{-3}$) = 2.25; log \textit{R (cm)}\ = 18.7 combinations (the same degeneracy occurs for other densities which have overlapping distances).  The average velocity of the 20 \Htwo\ absorption lines  discussed in \citet{abel16} is 19.5$\pm$0.7 \kms , clearly indicating that they arise in component III(B) The deduced physical conditions of Veil Component III (B) is independent of the best model for the \oi\ layer, with the exception that the \oi\ layer cannot be more than 10\textsuperscript{18.9} cm away, or else Component III (B) will produce more H\textsubscript{2}\ than observed.  \par
\begin{figure}
	\includegraphics
	[width=\columnwidth]
	{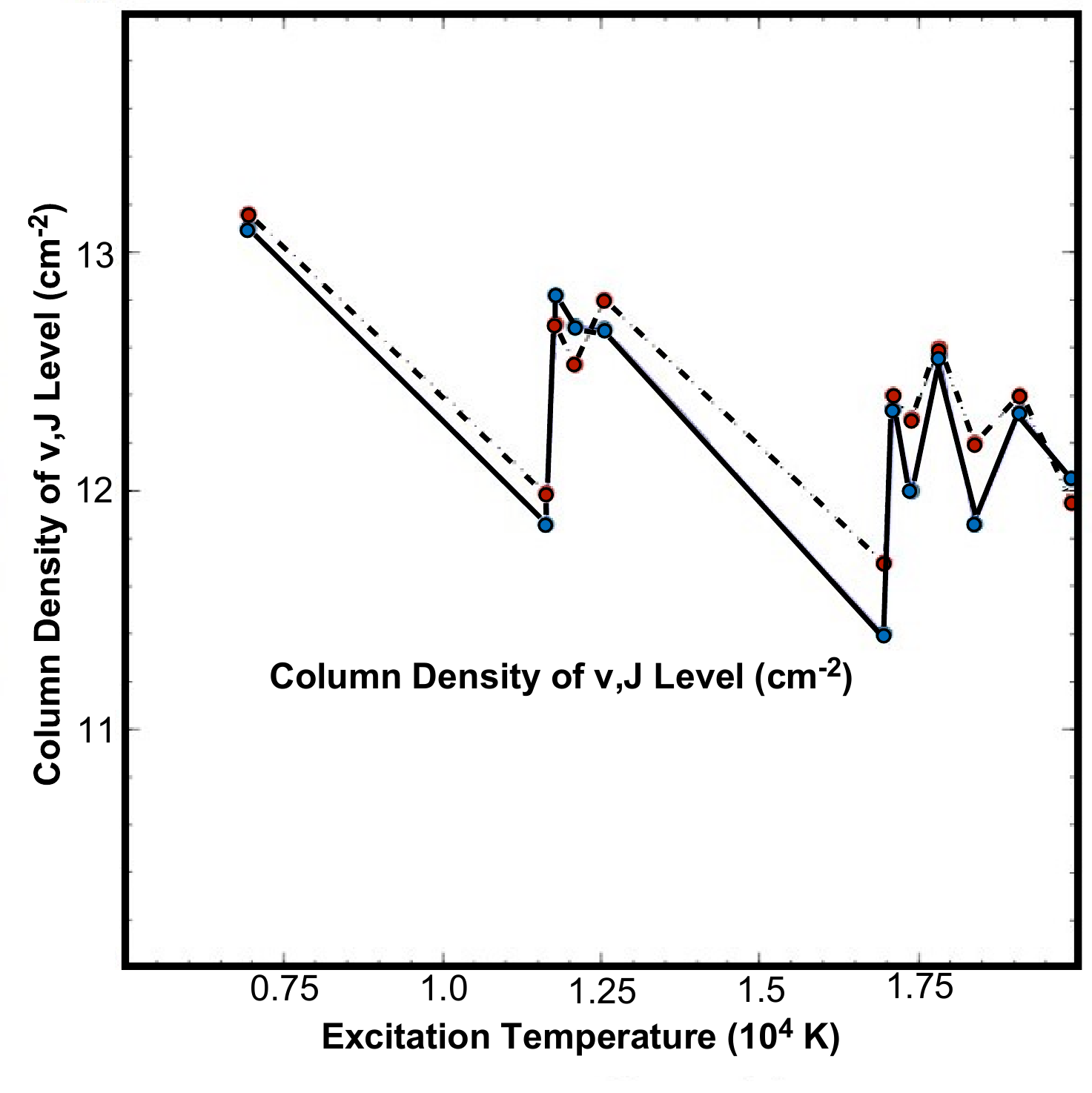}
\caption{Predicted (red) \Htwo\ column densities versus excitation temperature, compared to observation (blue), for the optimal density-distance combination deduced from the \oi , \nii , and \oiii , and HeI* observations.  Observed \Htwo\ (v,J) column densities are deduced from STIS UV absorption lines  analyzed in \citet{abel16}, while predicted \Htwo\ column densities come from the model \Htwo\ molecule incorporated in Cloudy  \citet{fer17}.  All \Htwo\ observations shown here correspond to the same velocity as Veil Component III (B),
 with v = 1 (J = 3), v = 2 (J = 0-3), and v = 3 (J = 0-6).}
\label{fig:H2}
\end{figure}
\vspace{\baselineskip}
The combination of distance-density that best reproduces Component III (B) is around a density of log \textit{n} (cm$^{-3}$) = 3 -- 3.2, and log \textit{R (cm)}\ = 18.8, which is nearly identical to previous modeling efforts of 
Component III (B) in \citet{abel06,abel16}. Figure~\ref{fig:H2} shows a comparison of the predicted \Htwo\ spectrum (red) for a density of log \textit{n} (cm$^{-3}$) = 3, log \textit{R (cm)} = 18.8 to the derived H\textsubscript{2}\ 
column densities from STIS observations (blue). Other calculations, such as CI, CI$\ast$,  and CI$\ast$ $\ast$, are unchanged from the published contour plots from \citet{abel16}.  

\subsection{Component IV (A)}
Given the results of calculations for Veil Component III (B), and the similarities between the \citet{abel16} paper and the current work, calculations for Component IV (A) are unnecessary, as given the same density, distance, and SED, Component IV (A) will produce the same deduced physical conditions as in the 2016 paper.  In \citet{abel16}, we deduced a density of log \textit{n} (cm$^{-3}$) = 2.4, and a distance of 4.2 parsecs.  As the density/distance controls other deduced physical conditions, such as the balance between magnetic, thermal, and turbulent energies, these deduced properties are also unchanged from  \citet{abel16}. \par

\section{The effects of \tA\ on our models}
\label{sec:tA}

\subsection{Observational Constraints}
\label{sec:tAobsConst}

In our models of the several components of the Veil we have only considered contributions of the Trapezium stars, dominated in the EUV by \tC, with FUV contributions by the later spectral type cooler
members of the Trapezium. The models presented do not assume any contribution by the O9.5V star \tA, but it is worth examining the effects of that star, which has much less EUV luminosity than \tC, but relatively more FUV luminosity. In order to be important in photoionizing Component I, \tA\ would have to lie on the side of the Veil away from the observer, although this location does not guarantee that it is important.

It is hard to determine where \tA\ lies along the line of sight relative to \tC. The first constraint on this comes from its placement relative to the Veil. 
The visual extinction of the Trapezium stars is about 
1.55  and that of \tA\ 0.7 \citep{gou82}, i.e. \tA\ has much less extinction. Since the extinction largely arises from within the Veil \citep{ode92b}, this appears to argue that \tA\ lies nearer to the observer than the Veil. However, the \hi\ 21 cm
absorption map, which shows the Veil to be thinning to the SW of the Dark Bay \citep{ode00}, which could account for the lower extinction of \tA\ even if it lies on the far side of the Veil. The \citet{ode93} study of very high
resolution spectra of HeI gives total column densities of the two (-2.9 \kms\ and 5.2 \kms) velocity components of
\tA\ of 2.94$\times$10$^{13}$~\cmsq, (individual values 1.70$\times$10$^{13}$~\cmsq and 1.24$\times$10$^{13}$~\cmsq respectively) and for \tC\ 1.45$\times$10$^{13}$~\cmsq.
 The average of the four Trapezium stars is 1.60$\times$10$^{13}$$\pm$0.34$\times$10$^{13}$~\cmsq. These numbers indicate that the Trapezium HeI components are very similar to those of the redder component in \tA. This suggest, but does not argue strongly for \tA\ being on the far side. However, the column density in \tC's HeI is high for the amount of extinction, when compared with the Trapezium stars, which may be due to different conditions of ionization.
 
 In a recent study \citet{ode17} of the ionization across the Huygens Region, they concluded that most of the MIF ionization northwest of the Bright Bar was caused by \tC, and that to the southeast by \tA. 
 However, they found that proplyds near the Bright Bar were ionized by \tA\  and some as close as half the distance in the plane of the sky were ionized by both \tA\ and \tC\ (one sees 
 ionization fronts in the same proplyd facing both stars ). Since the emission line proplyds clearly lie within the cavity between the MIF and the Veil, these objects become a strong argument that \tA\ lies on the far side of the Veil and its radiation can affect our Component I.
 It is prudent to investigate the sensitivity of our modeling efforts to \tA.
 
 \subsection{Models considering both \tC\ and \tA}
 \label{sec:tCandtA}

We used the best-fitting model to Component I (R (cm)= 10$^{18.6}$, n (cm$^{-3}$) = 10$^{2.65}$) and reran the model using an SED consistent with \tA.  We assume \tA\ is a typical O9.5V star with a stellar temperature of 34,600K and Q(H) = 3.63$\times$10$^{48}$ photons s$^{-1}$ \citep{ode17}.  Since \tA\ is about 0.41 parsecs away from the Trapezium in the plane of the sky, the distance of \tA\ from the line of sight to the Trapezium is highly uncertain.  We assume \tA\ is between the ionized layer and the MIF (just like the Trapezium).  Given this assumption, the closest distance \tA\ could be from the line of sight connecting the Earth to the Trapezium is 0.41 parsecs (if \tA\ were touching Component I).  If \tA\ were the same distance from Component I as the Trapezium (1.0 parsecs, for the best model), then \tA, the Trapezium, and the line of sight would form a right angle, and the \tA\ distance to the line of sight would be 1.08 parsecs.  

In addition to the distance, the direction the ionizing flux impinges Component I must also be considered.  Since radiation from \tA\ strikes the ionized layer at an angle (relative to the straight line connecting Trapezium to Earth), the flux will be reduced by the cosine of the angle between the Trapezium-Earth line of sight and a line connecting \tA\  to the illuminated face of Component I directly in front of the Trapezium.  We therefore calculate the ionizing photon flux, $\phi$(H) = Q(H)/4$\pi$R$^{2}$, where R$^{2}$ = 0.41$^{2}$ + y$^{2}$ is the distance of \tA\  from the illumination point and y is the perpendicular distance of \tA\  to Component I, allowed to vary between 0 and 1 parsec. We then multiplied by the ionizing flux by cos($\theta$), where cos($\theta$) = y/R.  Given these geometric assumptions, the minimum and maximum values for the contribution of \tA\ to $\phi$(H) comes out to 10$^{10.33}$ and 10$^{10.83}$ photons cm$^{-2}$ s$^{-1}$.  

Given the range of $\phi$(H) considered, we found a small, but ultimately inconsequential change to the model results for Component I.  Including \tA\ reduced the [N II] surface brightness from S([N II]) = 3.97x10\textsuperscript{-3} ergs s$^{-1}$ cm$^{-2}$ sr$^{-1}$ (without \tA) to (1.8-2.7)x10\textsuperscript{-3}ergs s$^{-1}$ cm$^{-2}$ sr$^{-1}$.  These slightly smaller values are still well within the observed range.  The [O III] surface brightness is identical to before, to within one percent, and the HeI* column density changed by less than 0.01 dex.  Overall, the best model for Component I is not significantly affected by including ionizing flux from \tA.  Since the model for Component II (the [O I] emitting layer) is based on the results of Component I, both ionized layers are relatively insensitive to \tA.

One way in which the presence of \tA\ could affect the Veil is the assumed distance of the neutral \hi\ layers (Component III (B) and IV (A)).  Including \tA\  increases the FUV flux, G\textsubscript{0}, by a factor of 2 - 4.  Increasing G\textsubscript{0} means we would need to increase the distance of Component III and IV by a factor of 1.4 to 2 in order to reproduce the same physical conditions deduced from UV and radio observations \citep{abel16}.  Overall, these results are expected, given the spectral type of \tA\ and we would expect the star to be a relatively minor contributor to the overall hydrogen ionizing flux, but perhaps a more important contributor to the FUV. 

 \section{Properties of a Central Column passing from the observer to the Orion Molecular Cloud}
\label{sec:what}

In Table~\ref{table:velocities} we present the velocities for multiple velocity components as derived in Section~\ref{sec:VelSys} and other studies cited in column 3. The assignment to components is by similar velocities
and expected states of ionization or illumination by \tC. The Backscattered MIF component is not a physical layer, rather, it is the observed velocity of MIF emission scattered in the PDR. We refer to this collection as representative of a Central Column because multiple stars are used for the absorption lines and the sampled emission line regions vary slightly between different studies. 

\subsection{Velocity components from \citet{goi15}}

The study of \citet{goi15} produced velocity images in CO 2-1, H41$\alpha$, and \Cii\ 158 $\mu$m. The data relevant to the spatial 
column discussed in this section are nicely presented in the Trapezium panel of velocities in their Figure~3. There one sees that the H41$\alpha$ line is wide, as caused by thermal broadening, and is centered on 16 \kms\ (2.5 \kms\ LSR). 
They assign this line to the MIF. 
The much narrower CO 2-1 line peaks at  28 \kms\ (10 \kms\ LSR) and is assigned to the background PDR and OMC.
The strongest \Cii\ 158 $\mu$m line peaks at 28 \kms\ (10 \kms\ LSR) and is assigned to the MIF.  

There is a weaker \Cii\ 158 $\mu$m line component at 19 \kms\ (0.5 \kms\ LSR). This velocity component shows up on their spectrum for the entire Huygens Region as a distinct shoulder on the blue side of the 28 \kms\ line. In Section 3.3.1 of \citet{goi15} those authors assign this weaker blue \Cii\ 158 $\mu$m component to a foreground layer that is illuminated by \tC 's strong FUV photodissociating continuum, based on of the absence of a similar CO 2-1 component, linking it to the \hi\ foreground layers. We assign it to Component III(B) because of the good agreement in velocity with the other members of this component. 

The detection of \Cii\ 158 $\mu$m emission is consistent with the observed UV absorption spectra and our theoretical modeling of the neutral Veil layers.  \citet{sof04} detected the C$^{+}$ 232.5 nm line in absorption towards the Trapezium, deducing a total column density along the line of sight of about 10$^{17.8}$ cm$^{-2}$.  The velocity of C$^{+}$ corresponds with the velocity of Veil Component III (B) and IV (A), indicating the observed C$^{+}$ is associated with the neutral layers along the line of sight.  Given the total H I column density towards the Trapezium (10$^{21.68}$ cm$^{-2}$, the ratio of ionized carbon to hydrogen is about 1.4x10$^{-4}$, which is consistent with the observed ratio of C/H in the ISM.  Therefore, almost all carbon along this sightline is in the form of C$^{+}$, and is associated with the neutral Veil layers.  The models computed in previous studies of the Veil (\citet{abel16} and references therein) assumes this ratio of C/H and stopped the calculation at the observed HI column density.

\subsection{Possible velocity components associated with a central shocked bubble created by \tC's stellar wind}
\label{sec:wind}

The \Vblue\ and \Vnew\  components are assigned to the boundaries of a 'Central Bubble' of hot shocked gas created by the strong stellar wind from \tC. The boundary closer to the observer is the \Vblue\ component and the further boundary is the \Vnew\ component. This \Vblue\ component is moving away from \tC\ at about 34$\pm$6 \kms\ , while the \Vnew\ component on the far side is moving away from \tC\ by 10$\pm$2 \kms\ for \nii\ and is almost stationary with respect to \tC\ in \oiii . The lower velocities in the far side of the Central Bubble are probably due to the shocked gas moving into higher density gas.  The boundaries in the plane of the sky are probably outlined by the 
High Ionization Arc that encircles the Trapezium region except in the direction of the Orion-S Cloud \citep{ode09b,ode18}. 

\begin{deluxetable*}{ccc}
\tablenum{2}
 \tablecaption{Velocity Systems*}
 \label{table:velocities}
\tablehead{
\colhead{Designation} &
\colhead{Component and Velocity* (\kms) } &
\colhead{Source}}
\startdata
 Veil Component IV(A)& \hi\ (absorption, 23.4)                   & \citet{vdw89,abel06}\\
 ---               & \Caii **(absorption, 22.0)              &\citet{ode93}\\
 ---               & \Nai ***(absorption, 23.0)               & \citet{ode93}\\
 Veil Component III(B) & \hi\ (absorption, 19.4)                & \citet{vdw89}\\
 ---& \Htwo\ (absorption 19.5$\pm$0.7) & \citet{abel16}\\
 ---& \Cii\ (19$\pm$2                              &\citet{goi15}\\
 ---                & \Caii ** (absorption, 18.3)            &\citet{ode93}\\
 ---                & \Nai *** (absorption, 19.8)             &\citet{ode93}\\
Neutral Oxygen Component II & \Vlowoi\ (11.0$\pm$1.5)  &\citet{jack00,ode18}\\
Ionized Component I& \Vlownii\ (5.8$\pm$3.8)             & NE--Region\\
 ---               & \Vlowoiii\ (7.8$\pm$2.1)            & NE--Region\\
 ---              & \oii\ (system B, 3.1)                    & \citet{jones92}\\
 ---                & \siii\ (system B, 9.9)                     & \cite{wen93}\\
 ---                & \heI\ (absorption, 2.1$\pm$0.6)&\citet{ode93}\\
 ---                & \Piii\ (absorption, 4.9$\pm$3.0) & \citet{abel06}\\
 ---                & \siiia\ (absorption,4.5$\pm$0.9)  & \citet{abel06}\\
 ---                 & \Caii ** (absorption, 7.5)              &\citet{ode93}\\
 ---                 & \Nai *** (absorption, 6.0)               & \citet{ode93}\\
Nearer Central Bubble     & \Vbluenii +\Vblueoiii\ (-7$\pm$6) & NE-Region\\
 Cluster Stars & Stellar Spectra (25$\pm$2)         & \cite{sa05}\\
 Further Central Bubble      & \Vnewnii\ (37$\pm$2)            & NE--Region\\
 ----                 & \Vnewoiii\ (27$\pm$5)           & NE--Region\\ 
 Main Ionization Front (MIF)& \oi\ (27$\pm$2)    & \citet{ode92a}\\
 ---                 &  \oii\ (18$\pm$1)           &\citet{adams44,jones92}\\
 ---               &  \nii\ (22$\pm$3)                  &NE--Region\\
 ---                 & \siii\ (20$\pm$4)                    &\cite{wen93}\\
 ---                 &  \oiii\ (18$\pm$3)                  &NE--Region\\
 ---                 & \Hplus\ (17$\pm$2)                &\citet{ode93}\\
 ---                 & H41$\alpha$ (16$\pm$2)        &\citet{goi15}\\
 ---                 & \Heplus\ (20$\pm$2)                & \citet{ode93}\\
 Photon-Dominated Region              & OMC molecules & \citet{ber14}\\
---                 &\Cii\ (28$\pm$2)               & \citet{goi15}\\
 Backscattered MIF & \nii\ (40$\pm$3)                & NE-Region\\
 ---                  & \siii\ (36$\pm$4)                 & \citet{wen93}\\
 ---                  & \oiii\ (38$\pm$3)                 & NE-Region\\
\enddata
*All velocities are Heliocentric velocities in \kms. They may be converted to LSR velocities by subtracting 18.1 \kms.

**\Caii\ 393.4 nm.
***\Nai\  589.6 nm.
\end{deluxetable*}

\subsection{Relative Velocities}
\label{sec:RelVel}

  \begin{figure*}
  \plotone{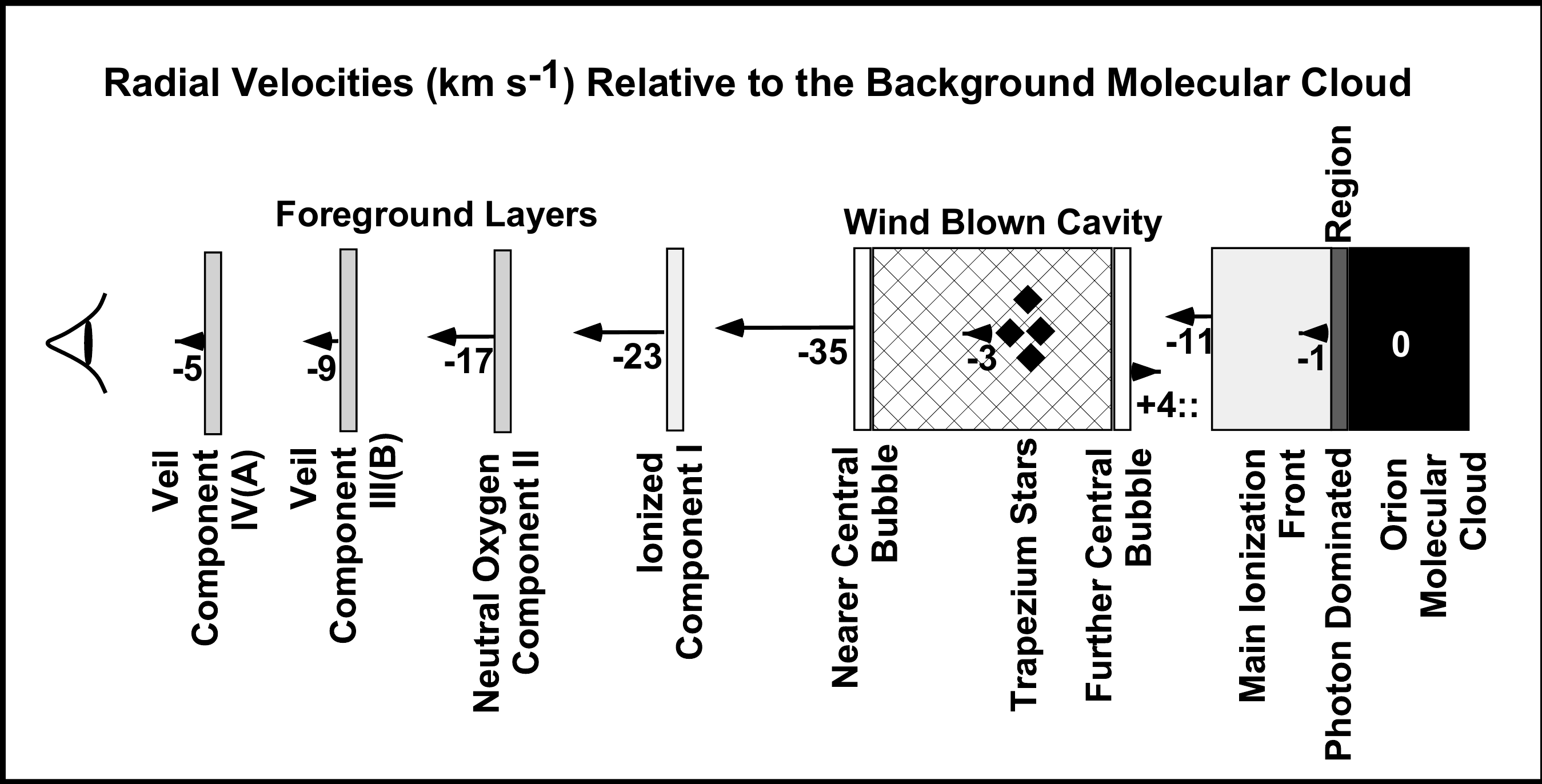}
    \caption{Like Figure~\ref{fig:Geometry} except now all layers in the direction of the central Huygens Region on both sides of \tC\ are shown.
    The velocity relative to the OMC velocity (28 \kms) is shown. For simplicity, 17 \kms\ is adopted for the MIF (Table~\ref{table:velocities} shows the Heliocentric Velocities for the various ionization layers within the MIF). Distances along the line-of-sight are not to scale, but they are indicated in Table~\ref{table:distances}. }
\label{fig:Layers}
\end{figure*}

Consideration of the relative velocities of the different velocity component sets the stage for relating this Central Column
to the over-all structure near the Huygens region. We'll use as the zero point the most massive component, the OMC.
The results are shown in Figure~\ref{fig:Layers}. The relative velocity of the MIF components (shown as -11 \kms ) is 
fully consistent with evaporation of material away the PDR (whose -1 \kms\ value is probably uncertain to 1 \kms). 
The relative velocities of the boundaries of the Wind Blown Cavity are consistent with expectation, with the expansion 
velocity on further side having been slowed by interaction with the dense MIF material and the nearer side showing a much higher relative velocity. This large velocity approaching velocity is contrary to the expectation that this inner cavity is
essentially stationary.

It is not surprising that the well defined foreground Veil components all have negative velocities with respect to 
the OMC because the ionization of the immediate environment of \tC\ and the stellar wind from that star will provide an outward push. However, Components III(B) and IV(A) may have velocities inherently different because of their distance from the OMC.  As noted before \citep{abel16}, the negative velocity differences mean that Components I and II are going to collide with component III(B) at a time determined by the separation of the layers. A detailed study of this collision
timescale may tell us about the lifetime of the stellar wind and FUV flux of \tC.

\subsubsection{A possible problematic correlation of \Vlow\ and \Vmif\ velocities}
\label{sec:VlowVmif}

O'Dell's (2018) recent study has shown that there is a good correlation of \nii\ and \oiii\ Component I 
and MIF velocities if one considers both large samples and individual slit data of lower accuracy, which is difficult to understand, given their separation.  The correlation is not clear when only considering 10\arcsec$\times$10\arcsec~samples. In any event there  is a well defined velocity difference of -18 \kms\ for \nii\ and a less well defined 
velocity difference of -13 \kms\ for \oiii\ over a \Vmif\ range of 10 \kms. Our NE samples (Figure~\ref{fig:Histograms}) give differences of -17$\pm$6 \kms\ for \nii\ and -10$\pm$2 \kms\ for \oiii. 
How a line of site correlation can exist presents a puzzle since Component I lies about 1.3 pc in the foreground of \tC\ 
and the MIF about one-fourth pc beyond \tC .

Could the X-ray emitting hot gas discovered by \citet{gud08} affect Component I along with the other Veil layers?  In planetary nebulae, where more complete observations exist, the hot and nebular phases are in rough pressure equilibrium.  There are no observations of the hot gas for the regions close to the Trapezium that we study here due to extinction of soft X-rays by the Veil.  \citet{art12}  made hydrodynamical simulations of the hot gas with the aim of matching the G\"udel observations.  Her Figure 2 suggests a gas pressure of $nT \sim 10^6$ K cm$^{-3}$ close to the stars.  This is very uncertain since the calculations were not intended to reproduce the properties of the hot gas close to the star cluster, never-the-less they do serve as a rough guide to what might be present.  This pressure is similar to the gas pressure in the two Veil components (Component I $10^{6.5}$ K cm$^{-3}$ and Component 2 $10^{6.0}$ K cm$^{-3}$).  It is possible that the Veil layers we observe are entrained with the hot gas flow.  By comparison, the MIF has a much higher gas pressure, $10^{7.5}$ K cm$^{-3}$, so would not be affected by the pressure from the X-ray gas.  The MIF pressure would push on the hot bubble, which could in turn push the foreground Veil layers.  This would account for the large scale correlation between the two velocities. This process would not explain a \Vmif\ and \Vlow\  correlation on a small scale in the plane of the sky.

\section{How do these results relate to the structure of the EON?}
\label{sec:EON}
\begin{deluxetable}{lc}
\tablenum{3}
 \tablecaption{Distance from \tC} \label{table:distances}
\tablehead{
\colhead{Feature} &
\colhead{Distance}}
\startdata
Veil Component IV(A)& 4.2* pc\\
Veil Component III(B) & 2.0* pc\\
Veil Component II &    $\geq$1.3* pc, $\leq$2.0* pc\\
Veil Component I  & 1.3 pc\\
Nearer Central Bubble & 0.1** pc\\
\tC & 0 \\
Photon-Dominated Region & -0.3* pc\\
\enddata
*Derived from Ionization Modeling.

**One half of apparent diameter of the High Ionization Arc.
\end{deluxetable}

In Table~\ref{table:distances} we summarize the distances of the various major components within our Central Column.
Most of these are derived from predictions from ionization modeling with observations. The exception to this is the Nearer Central Bubble value of 0.1pc. In this case we have adopted one-half of the diameter of the High Ionization Arc (106\arcsec\ or 0.20 pc diameter). 

For reference, we note that the north-south axis of the EON optical window image is about 34\farcm7\ (3.9 pc) and the east-west axis is about 30\farcm1\ (3.3 pc). If the EON is one-half of a spherical bubble, its near side would be about 1.8 pc in front of \tC , a value very near the distance of Veil Component III(B). However, there is little to support the assumption of that geometry and \tC\ is not near the center of the EON, lying only 0.4 pc from the north-east corner of the EON. 

A very relevant recent paper \citet{pabst} examines the EON structure using \Cii\ 158 $\mu$m high velocity resolution images obtained with the airborne observatory SOFIA. That paper does not use the term EON, rather, they call it the Veil Bubble because of the interpretation they apply to their important new observations. Of particular interest is their conclusion 
that the EON actually extends as far as 2500\arcsec , which corresponds to a diameter of 4.6 pc at the distance we adopt 
(383$\pm$3 pc, \citep{mk17,gro19} or 5.0 pc at the distance (414$\pm$7, \citep{men07}) they adopt.  Position-velocity \Cii\ 158 $\mu$m emission plots across the EON (both north-south and east-west) indicate an irregular and weak \Cii\ 158 $\mu$m component with a velocity of expansion of 13 \kms\ 
away from the MIF component. In the vicinity of the Huygens region this component is displaced about 8 \kms , in agreement with our Central Column \Cii\ 158 $\mu$m displacement of 9 \kms\ derived from the \citet{goi15} study. Their interpretation is that the displaced \Cii\ 158 $\mu$m component is the boundary of a 2 pc radius shell encompassing a high temperature region that is the source of the X-ray emission. A smaller region (which they do not detect but include in their model) contains a 
sphere of gas evacuated by the fast stellar wind of \tC. They do not use the 21 cm \hi\  data because of the complexity of the emission and absorption results in this velocity region. 

In our study of the Central Column near \tC\ we show strong evidence for linking the expanding \Cii\ 158 $\mu$m material to Component III(B), which is seen in \hi , \Htwo, \Caii , and \Nai. In addition, we establish that Component IV(A) 
must lie outside of Component III(B) and that the well defined Component I must lie between both Component III(B)
and the ionized Component II. We also argue that the High Ionization Arc is the edge of the sphere of low density 
material swept up by the stellar wind from \tC. This not to argue that the \citet{pabst} model is wrong, rather, that the real 
situation is more complex than pictured from only the \Cii\ 158 $\mu$m line and complementary infrared data. The evidence for Component III(B) being part of a curved shell of about the same spatial size as the EON is strong, while we cannot comment on the form of Component I, Component II, and Component IV(A).

\section{Conclusions}
\label{sec:conclusions}
The foreground Veil associated with the Huygens Region of the Orion Nebula has recently been shown to have multiple 
new components. 

Comparison of the observed and calculated surface brightnesses for \nii , \oiii, and \oi\ indicate the conditions within the two Veil layers closest to the Trapezium stars. The strength of the \heI\ absorption lines in the Trapezium stars allow further refinement of the conditions within the Veil.

We establish that there are two faint but real
ionized layers on either side of Trapezium and these probably represent the two sides of the hot central cavity caused 
by the strong stellar wind from \tC. The high pressure of this bubble is the driving force that causes the Veil components
to be moving with a blueshift relative to the host Orion Molecular Cloud. This also provides the explanation for the radial velocities of the ionized Veil layer closest to \tC\ being correlated to the velocities of the MIF.

We calculated the effects of \tA\ on the deduced physical conditions, and find a negligible contribution to the ionized layers, but the FUV contribution to the overall SED could lead to a factor of two increase in the distance of the neutral layers from the Trapezium.

We confirm that one layer (Component III(B)) in our Central Column is 
probably a part of a recently recognized expanding shell of material covering the entire EON.

 \section*{acknowledgements}
In this study we have made extensive use of the SIMBAD database, operated at CDS, Strasbourg, France and its mirror site at Harvard University and to NASA's Astrophysics Data System Bibliographic Services. We have used IRAF, which is distributed by the National Optical Astronomy Observatories, which is operated by the Association of Universities for Research in Astronomy, Inc.\ under cooperative agreement with the National Science foundation. 

The observational data were obtained from observations with the NASA/ESA Hubble Space Telescope,
obtained at the Space Telescope Science Institute, which is operated by
the Association of Universities for Research in Astronomy, Inc., under
NASA Contract No. NAS 5-26555; the Kitt Peak National Observatory and the Cerro Tololo Interamerican Observatory operated by the Association of Universities for Research in Astronomy, Inc., under cooperative agreement with the National Science Foundation; and the San Pedro M\'artir Observatory operated by the Universidad Nacional Aut\'onoma de M\'exico. The MUSE observations at Paranal were made under the auspices 
of the European Southern Observatory. 

GJF acknowledges support by NSF (1816537), NASA (ATP 17-ATP17-0141), and STScI (HST-AR- 15018).

The authors are grateful to the anonymous referee, whose comments led to significant improvements in our investigation.

\clearpage

\end{document}